\documentstyle[preprint,floats,aps,psfig,draft]{revtex}
\tightenlines
\begin{document}

\draft

\title{
Statistical Mechanics of Soft Margin Classifiers
}

\author{Sebastian Risau-Gusman\dag\ddag \, and
Mirta B. Gordon\dag \cite{cnrs}\\
\dag\ D\'epartement de Recherche Fondamentale sur
la Mati\`ere Condens\'ee\\ CEA - Grenoble, 17 rue des Martyrs,
38054 Grenoble Cedex 9, France\\
\ddag\ Zentrum f\"ur interdisziplin\"are Forschung \\
Wellenberg 1, D-33615 Bielefeld, Germany
}

\date{\today}

\maketitle

\begin{center}
\begin{abstract}
We study the typical learning properties of the recently
introduced Soft Margin Classifiers (SMCs), learning realizable
and unrealizable tasks, with the tools of Statistical
Mechanics. We derive analytically the behaviour of the
learning curves in the regime of very large training
sets. We obtain exponential and power laws for the decay of
the generalization error towards the asymptotic value,
depending on the task and on general characteristics of
the distribution of stabilities of the patterns to be
learned. The optimal learning curves of the SMCs, which
give the minimal generalization error, are obtained by
tuning the coefficient controlling the trade-off between
the error and the regularization terms in the cost function.
If the task is realizable by the SMC, the optimal performance
is better than that of a hard margin Support Vector Machine and is very close to that of a Bayesian classifier.
\end{abstract}

\end{center}


\section{Introduction}

Neural networks are models of learning systems composed of
interconnected units that, besides their biological relevance,
have been shown to be very useful for classification tasks. The
weights of the connections are adjusted through a process called
{\it learning} using a set of $M$ examples. It is assumed that
these are labeled following an underlying rule, usually called
{\it teacher}. The purpose of learning is not only to classify
correctly the examples of the training set, but also to {\it
generalize} correctly on new inputs. To this aim, the network has
to infer the teacher's rule. The quality of this inference is
measured through the generalization error $\epsilon_g$, which is
the probability of misclassification of a new, randomly selected,
input pattern. As $\epsilon_g$ is not a quantity available for the
training process, learning is usually performed through the
minimization of a function of the training patterns. The tools of
Statistical Mechanics allow to study the properties of such
learning systems, providing a deep understanding of their typical
behaviour~\cite{Gardner,GyTi,HKP,SST,Peretto}. In particular, it
has been shown that the minimization of the training error, that
is, the fraction of training patterns misclassified by the
network, does not necessarily provide the best
generalizer~\cite{MeFo2,KiCa2,BuToGo}. This is why other cost
functions, based on geometrical properties like the distance of
the patterns to the discriminating surface, or on probabilistic
error measures like the likelihood, are used for training.

The simplest instance of a neural network, the
perceptron, is a single binary unit whose output
is the sign of the weighted sum of its inputs. It
can only perform linear separations of the patterns.
If the classification task requires more complex
discriminating surfaces, these may be implemented
using feedforward networks with a layer of hidden
units whose number is a priori unknown. The cost functions
used to tackle this problem usually have several minima,
and determining the lowest one is one of the main
difficulties of learning with multilayer neural networks.
This is also a problem for the theoretical analysis, as the
typical properties of such networks depend crucially on the
structure of the minima in the weights' space.

Recently, a new learning scheme has been proposed, which
strives to get rid of the problem raised by the multiple
minima. The obtained classifiers are called {\it Support
Vector Machines} (SVMs)~\cite{Vapnik,CoVa}. Instead of
directly looking for a complicated discriminating surface
in input space, the patterns are first mapped to a high
dimensional {\it feature space}, where the rule to be
learned is (hopefully) linearly separable. If this is the
case, a simple perceptron can be trained to find the
separation in feature space. Denoting the weights by
${\bf w} \in \Re^N$, the perceptron's output to an input
${\bf x} \in \Re^N$ is given by $\sigma={\rm sign}({\bf w} \cdot
{\bf x} + b)$ where $b$ is a bias and the dot represents
the inner product in $\Re^N$. Thus, the patterns belonging
to different classes are separated by a hyperplane orthogonal
to ${\bf w}$ at distance $|b|/\|{\bf w}\|$ from the origin, with $\|{\bf w}\|=\sqrt{{\bf w}\cdot{\bf w}}$. The SVM's
solution is the {\it Maximal Stability Perceptron}
(MSP)~\cite{KrMe2} in feature space, also called maximal
margin hyperplane. This is the hyperplane at maximal
distance $\kappa_{max}$ from the closest patterns in the
training set. Two different formulations of this problem
in terms of cost functions have been proposed in the literature.
In the first one~\cite{KrMe2}, the cost function counts
not only the number of misclassified patterns, but also the
number of correctly classified ones that lie at a distance
smaller than $\kappa$ from the separating hyperplane:

\begin{equation}
\label{eq.MSP}
E_{\rm MSP}(\kappa) = \sum_{\mu=1}^M \Theta(\kappa \|{\bf w}\|-h_\mu),
\end{equation}

\noindent where $\Theta$ is the Heaviside function, and

\begin{equation}
\label{eq.aligned}
h_\mu \equiv \tau_\mu ({\bf w} \cdot {\bf x}_\mu+b),
\end{equation}

\noindent is called {\it aligned field} of the training
pattern ${\bf x}_\mu$, $\tau_\mu \in \{-1,1\}$ being its
class. If
the $M$ $N$-dimensional patterns are correctly classified,
the aligned fields are all positive. The SVM solution has
$\bf w$ and $b$ corresponding to $\kappa_{max}$, the largest
possible value of $\kappa$ such that $E_{\rm MSP}(\kappa_{max})=0$.
If the training set is not linearly separable, $\kappa_{max}$
becomes negative. Notice that there are no constraints on
the norm of $\bf w$, that can be freely chosen.

If the norm of the weight vector is chosen so that the
aligned field of the closest pattern be $1$, this
leads to an equivalent formulation of the problem~\cite{Vapnik,CoVa},
in which the function to be minimized is:

\begin{equation}
\label{eq.costprimMSP}
E_{\rm SVM} = \frac{1}{2} {\bf w} \cdot {\bf w},
\end{equation}

\noindent subject to the conditions

\begin{equation}
\label{eq.condprimMSP}
h_\mu \geq 1,\,\,\,\,   \mu=1,...,M.
\end{equation}

\noindent Clearly, the constraints (\ref{eq.condprimMSP})
can only be satisfied if it is possible to classify
correctly all the examples. In that case, there are no
training patterns in a strip of width $1/\|{\bf w}\|$ on
both sides of the hyperplane, meaning that in the error-free
regime $1/\|{\bf w}\| \equiv \kappa_{max}$. An interesting
property of the SVM solution is that the weight vector and
the bias can be written as a linear combination of a sub-set
of training patterns, the {\it Support Vectors}, having
$h_\mu=1$.

The minimization of (\ref{eq.MSP}) with $\kappa=\kappa_{max}$
is equivalent to that of (\ref{eq.costprimMSP}) with condition
(\ref{eq.condprimMSP}) {\it only if the training set
is linearly separable}. If errors cannot be avoided,
the equivalence breaks down, as in one hand (\ref{eq.MSP})
has either negative $\kappa_{max}$, or several minima
if $\kappa_{max} \geq 0$ is imposed, and on the other
hand the constraints (\ref{eq.condprimMSP}) cannot be
satisfied. This is why the second formulation has been
generalized~\cite{CoVa} through the introduction of a
new set of variables $\zeta_\mu \ge 0$, called
{\it slacks}, which are a measure of the ``amount of
violation'' of the constraints. An increasing
function of these is included in the cost function
(\ref{eq.costprimMSP}) and the hard margin conditions
(\ref{eq.condprimMSP}) are modified to allow some
patterns to be closer to the hyperplane than
$1/\|{\bf w}\|$. The new problem amounts to minimize:

\begin{equation}
\label{eq.costprimSMC}
E_{C,k} = \frac{1}{2} {\bf w} \cdot {\bf w}+C \sum_{\mu=1}^M
{\zeta_\mu}^k,
\end{equation}

\noindent subject to the following conditions for $\mu=1,...,M$

\begin{mathletters}
\begin{eqnarray}
\label{eq.condprimSMC1}
h_\mu &\geq& 1 - \zeta_\mu, \\
\label{eq.condprimSMC2}
\zeta_\mu &\geq& 0.
\end{eqnarray}
\end{mathletters}

\noindent The coefficient $C$ in (\ref{eq.costprimSMC}) is
a hyperparameter that allows to control the trade-off between
the error term, defined by the slacks, and the regularization
term, proportional to the squared weights. As will be shown in section
\ref{sec:Copt}, it may be selected to optimize the generalization
performance. The exponent $k$ in (\ref{eq.costprimSMC}) modulates
the relative cost of errors, depending on their distance to the
hyperplane. Patterns in a strip of width $1/\|{\bf w}\|$ at each
side of the hyperplane, whether correctly or incorrectly
classified, as well as those incorrectly classified
outside of this strip, have $\zeta_\mu >0$. $1/\|{\bf w}\|$
is called {\it soft margin}, and the resulting classifier
{\it soft margin} SVM or {\it soft margin classifier} (SMC).

As the cost (\ref{eq.costprimSMC}) is a quadratic function for $k=1$ and $k=2$,
and the domain of minimization defined
by (\ref{eq.condprimSMC1}) and (\ref{eq.condprimSMC2}) is
convex, the minimum is {\it unique}~\cite{BuCr} . This remarkable
property makes the new formulation attractive for
applications, as it allows to get rid of the multiple
minima appearing in other learning schemes. Like in the
hard margin formulation, the solution $\{{\bf w}, b\}$ can
be expressed as a linear combination of the {\it support vectors},
which now include the patterns with positive slacks. The corresponding
coefficients may be obtained by solving the {\it dual}
problem (see for example~\cite{Martos}) which, for
$k=1$ or $k=2$ has a particularly simple
expression~\cite{CoVa}. Several efficient methods are
known for solving this kind of problems, and this is one of
the reasons why these classifiers are so widely used lately.

In this paper we study the typical properties of the
SMCs obtained by solving equation (\ref{eq.costprimSMC})
subject to the conditions (\ref{eq.condprimSMC1}) and
(\ref{eq.condprimSMC2}), with the methods of Statistical
Mechanics, using the replica approach. It has been
shown~\cite{RGGo1,RGGo2} that the statistical properties
of SVMs in high dimensional feature spaces~\cite{DOS} can
be well approximated by considering a simple perceptron
learning anisotropically distributed patterns. The
amount of anisotropy depends on the normalization of the
mapping from the input to the feature space. In this
paper we restrict to an isotropic pattern distribution,
which corresponds to a non-normalized mapping.

The learning properties of a perceptron learning an
isotropic input pattern distribution have
been extensively studied~\cite{OpKi},
mainly for linearly separable, i.e. realizable,
tasks. In this case the hypothesis of replica symmetry
is generally correct, allowing for a full analytical
statistical mechanics calculation. In particular, the
behaviour of the generalization error $\epsilon_g$ in
the limit of very large $\alpha \equiv M/N$ has a
universal power law decay $\epsilon_g \approx \alpha^{-\nu}$
with $\nu=1$. Its prefactor allows to characterize the
convergence to perfect learning of different learning
algorithms. If the rule to be inferred cannot be
generalized without errors, the task is called
{\it unrealizable}. In this case  the replica symmetric
solution, although generally unstable, is believed
to provide a good approximation of some learning
properties. However, in the case of a linearly separable rule
learned with noisy training patterns, which is thus
unrealizable, the replica symmetric approximation gives an exponent $\nu=1/2$~\cite{GyTi} whereas one step of replica
symmetry breaking shows~\cite{UeKa} that this exponent
is modified to $\nu=2/3$. As this is but an approximation
to the full replica symmetry breaking scheme~\cite{ReGy}
at zero temperature, it is not clear whether this exponent
is correct. The same exponent has been found in the case
of a quadratic hard margin SVM learning a linearly separable
task, that is, a rule simpler than those implementable
with the student's architecture~\cite{DOS}. Another case
of interest is that of {\it inconsistent learning}~\cite{MeFo2},
which refers to realizable tasks learned with algorithms
that do not strive to minimize the number of training errors.
In this case, the exponent within the replica symmetric
approximation was found to be $\nu=1/2$~\cite{MeFo2}.

As the soft margin problem has a unique minimum for $k=1$ and $k=2$,
even if the task is unrealizable, the replica symmetry
hypothesis should be always correct, providing a framework
for the study of complex classification tasks even when the
mismatch between the student and the teacher hinders
error-free learning.

In this paper we present the statistical properties of
SMCs learning several kinds of realizable and unrealizable
rules. The model and the statistical mechanics approach
are presented in section \ref{sec:statmech}. The theoretical
properties of SMCs with exponents $k=1$ and $k=2$ in
the cost function (\ref{eq.costprimSMC}) are obtained as
a function of the training set size $\alpha \equiv M/N$ in
the thermodynamic limit $N, M \rightarrow \infty$. Several
teacher rules are considered in section \ref{sec:learning}.
One of our most striking results is that the generalization
error for large $\alpha$ exhibits a very rich variety of
asymptotic behaviours, depending on the type of rule to be
inferred. In particular, even if the task is realizable, the
soft margin algorithm is inconsistent unless $C \rightarrow
\infty$. For finite $C$, we find that the fraction of training
errors at finite $\alpha$ is finite, and the generalization
error vanishes asymptotically with $\alpha$ following a $\nu=2/3$
power law. In the unrealizable tasks considered, $\epsilon_g$
converges to an asymptotic finite value either exponentially
or with a power law with $\nu=1/2$ . The usual exponent $\nu=1$
only arises for error-free learning of a realizable task. In
section \ref{sec:Copt} we derive the best generalization
performances of SMCs through the determination of the value
$C_{opt}(\alpha)$ that minimizes the generalization error.
Finally we present a summary of our results in section
\ref{sec:Discussion}, together with some open questions.
Most details of the proofs are left to the Appendix.

\section{Statistical mechanics approach}
\label{sec:statmech}

We consider a student perceptron of weight
vector ${\bf w}=(w_1, \dots, w_N)$, {\it without threshold}.
That is, we set $b=0$ in (\ref{eq.condprimSMC1}). Given any
$N$-dimensional input vector $\bf x$, the classifier's output is
$\sigma={\rm sign}({\bf w} \cdot {\bf x})$: all the
points lying on the same side of a hyperplane orthogonal
to ${\bf w}$ containing the origin are given
the same class. We assume that the perceptron learns
the classification with the soft margin algorithm,
using a set ${\mathcal L}_M=\{ ({\bf x}_\mu,\tau_\mu) \}_{\mu=1, \dots,M}$
of $M$ examples or training patterns. These consist of input
vectors ${\bf x}_\mu$ drawn from an isotropic gaussian
distribution of variance $1/\sqrt{N}$,

\begin {equation}
\label{eq.probax}
P({\bf x})=\frac{e^{-N{\bf x}^2/2}}{(2 \pi/N)^{N/2}};
\end{equation}

\noindent and labels $\tau_\mu \in \{ -1,1\}$
that represent the corresponding classes. The classification
tasks considered in this paper are given by the following
teacher's rule:

\begin{equation}
\label{eq.rule}
\tau = {\rm sign}\left({\cal P}({\bf w}_0 \cdot {\bf x})\right),
\end{equation}

\noindent where ${\bf w}_0$ is referred to as the
teacher's vector hereafter, and ${\cal P}(z)$ is
a polynomial of $z$. Each of its zeros $z_i$~\cite{foot0}
defines a discriminating hyperplane
at a distance $|z_i|/\|{\bf w}_0\|$ from the origin.
Rules of the kind (\ref{eq.rule}) partition the input
space in as many different regions as the number of zeros
of the polynomial plus one, separated by parallel hyperplanes
normal to the teacher's vector ${\bf w}_0$. Patterns in
successive regions belong alternatively to class $+1$ or
$-1$. As only the zeros of the function ${\cal P}(z)$ matter,
there is no loss of generality in our assumption that
${\cal P}(z)$ is a polynomial. We assume $\|{\bf w}_0\|=\sqrt{N}$,
which is equivalent to imposing the unit of distance. Notice that
the only rule realizable for the student perceptron considered
in this paper is that of the linear teacher ${\cal P}(z)=z$.

In the following we study the properties of the
solution to the soft margin problem using the by
now standard tools of Statistical Mechanics~\cite{Gardner,GyTi}.
That is, we assume that the ensemble of classifiers
follows a Gibbs distribution defined by the energy
function (\ref{eq.costprimSMC}), at a fictitious
temperature $1/\beta$, and we take the zero
temperature limit. The constraints (\ref{eq.condprimSMC1})
and (\ref{eq.condprimSMC2}) play the role of infinite
potential walls. Notice that the phase space in the present
case has dimension $\Re^{N+M}$, as not only the weights
${\bf w}$ but also the slacks $\{{\bf \zeta}_\mu\}_{\mu=1,\dots,M}$,
have to be learned. The partition function is:

\begin{equation}
\label{eq.partition}
Z_{C,k}(\beta;{\mathcal L}_M,{\cal P}) =
\int \exp \left(-\beta E_{C,k}({\bf w},\{ {\bf \zeta_\mu}\}) \right) \;
\prod_{\mu=1}^{M} \Theta \left(\tau_\mu {\bf w} \cdot {\bf
x_\mu}-(1-\zeta_\mu) \right) \Theta(\zeta_\mu)\, d{\bf w} \,d{\bf \zeta_\mu}.
\end{equation}

The inverse temperature $\beta$ has obviously no physical
meaning whatsoever; it is only introduced in order to study
the properties of the SMC which, being the single minimum
of the energy function, is selected in the limit
$\beta \rightarrow \infty $. We assume that the number
of training examples scales with the input space dimension,
$M=\alpha N$, and take the thermodynamic limit
$N \rightarrow \infty$, $M \rightarrow \infty$ with
$\alpha \equiv M/N$ constant. The free energy
per input space dimension averaged over all the possible
training sets of $M$ patterns, $f_{C, k}(\beta;{\cal P})$,
is calculated with the replica method, that uses the identity

\begin{equation}
\label{eq.elibre}
f_{C, k}(\beta;{\cal P}) = -\lim_{N \rightarrow \infty} \frac{1}{N \beta} \, \overline {\ln \, Z_{C,k}(\beta;{\mathcal L}_M,{\cal P})}
=-\lim_{N \rightarrow \infty} \frac{1}{N \beta} \, \lim_{n \rightarrow 0}
\frac{\ln {\overline {Z^n_{C, k}(\beta;{\mathcal L}_M,{\cal P})} }}{n},
\end{equation}

\noindent where the overline represents the average over
the pattern distribution (\ref{eq.probax}), with labels
given by (\ref{eq.rule}). $Z^n$ is the
partition function of $n$ independent replicas of the
problem, that become coupled after taking the average.
The typical properties of the classifier are obtained
by taking the limit $\beta \rightarrow \infty$. The free
energy (\ref{eq.elibre}) turns out to be a function of
the following order parameters:

\begin{mathletters}
\begin{eqnarray}
\label{eq.Q}
Q_a  &=& \frac{\overline{\langle{\bf w}_a  \cdot  {\bf w}_a \rangle}}{N},
\\
\label{eq.q}
q_{ab} &=& \frac{\overline{\langle{\bf w}_a  \cdot  {\bf w}_b } \rangle}{N},
\\
\label{eq.R}
\tilde{R_a} &=& \frac{\overline{\langle{\bf w}_a  \cdot  {\bf w}_0 \rangle}}{N},
\end{eqnarray}
\end{mathletters}

\noindent where the brackets represent the phase space
average and $a$ and $b$ are replica indices. The norm
of the perceptron's weight vector, $Q_a$, is one of the order
parameters because in the soft margin problem the weights
are not normalized as usually. $q_{ab}$ is the overlap
between two different weight vectors at temperature
$\beta^{-1}$, and ${\tilde R}_a$ is the overlap of the
perceptron's solution and the teacher's vector.

As for $k=1$ and $k=2$ the energy in (\ref{eq.costprimSMC}) is
a quadratic function in a convex domain, it has a single
minimum~\cite{foot1}, irrespective of the kind of rule that
is being learned. Therefore, we may safely assume that all the
replicas are equivalent, even in the case of learning unrealizable
rules. We obtain thus the typical properties for cases where,
using other more usual cost functions like the number of training
errors, full replica symmetry breaking would be required~\cite{ReGy}.
The excellent agreement of the theoretical predictions and
the numerical simulations presented in the following section
is a further justification of our hypothesis of {\it replica
symmetry}. Thus, we set $Q_a \equiv Q$, $q_{ab} \equiv q$ and
$\tilde{R_a} \equiv \tilde R$, and we define the normalized
overlap $R \equiv \tilde{R}/\sqrt{Q}$, that only depends on
the angle between ${\bf w}$ and ${\bf w}_0$.

Due to the unicity of the soft margin solution, only one point
in phase space has non vanishing probability in the limit
$\beta \rightarrow \infty$, so that $q \rightarrow Q$. It is
convenient to introduce a new parameter, $x \equiv \beta(Q-q)$,
which reflects how fast the fluctuations around the minimum
of (\ref{eq.costprimSMC}) vanish as $\beta \rightarrow \infty$.
In this limit we obtain the typical free energy of the SMC learning a
rule defined by the polynomial ${\cal P}$,

\begin{equation}
\label{eq.f}
f_{C, k}({\cal P}) = -{\rm extr}_{\{Q,R,x\}}\,\left(G_0(Q,R,x) -\alpha G_{C,k}(Q,R,x;{\cal P})\right) ,
\end{equation}

\noindent where

\begin{equation}
\label{eq.G0}
G_0(Q,R,x)=\frac{Q}{2x} (1-R^2-x),
\end{equation}

\noindent is an entropic term. The dependence on the rule to
be learned is embodied in the second term of (\ref{eq.f}) through
${\cal P}(z)$, and on the learning algorithm through $k$ and $C$.
Integrating out the slack variables in the limit
$\beta \rightarrow \infty$ through a saddle point approximation,
we get

\begin{equation}
\label{eq.G1}
G_{C,k}(Q,R,x;{\cal P}) =
\int_{- \infty}^{\infty} Dy \int_{\phi(y;Q,R,{\cal P})}^{\infty} Dt \;
\min_{\zeta}  W(\zeta; y,t,Q,R,x,{\cal P}),
\end{equation}

\noindent where $Dt \equiv dt \, \exp{(-t^2/2)}/\sqrt{2 \pi}$,

\begin{equation}
\phi(y;Q,R,{\cal P})=\frac{y R \; {\rm sign}({\cal P}(y))-1/\sqrt{Q}}{\sqrt{1-R^2}},
\end{equation}

\noindent and

\begin{equation}
W(\zeta; y,t,Q,R,x,{\cal P})= C \,{\zeta}^k +
\frac{\left(\zeta-\sqrt{Q(1-R^2)}\,(t-\phi(y;Q,R,{\cal P}))\right)^2}{2x}.
\end{equation}

\noindent In (\ref{eq.G1}), due to the saddle point approximation,
$W(\zeta; y,t,Q,R,x,{\cal P})$ has to be taken at its minimum
$\zeta(t,y) \in [0,\sqrt{Q (1-R^2)} \phi(y;Q,R,{\cal P})]$ for
each couple $(y,t)$.
It is easy to see that there is a unique
local minimum inside this interval for $k>1$.
For $k=1$, $W$ is a quadratic function of $\zeta$, whose global
minimum falls inside the allowed interval only for a finite range
of values of $t$. Outside this range, the minimum lies at the
boundary $\zeta=0$. As a consequence, for $k=1$ the inner integral
in $G_{C,k}$ splits into two parts. The results for
$k=1$ and $k=2$ are respectively:

\begin{eqnarray}
\label{eq.G1k1}
G_{C,1}(Q,R,x;{\cal P}) &= &\int_{\frac{-1}{\sqrt{Q}}}^{\frac{xC-1}{\sqrt{Q}}} Dt \frac{(t \sqrt{Q} + 1)^2}{2x} \, g(t;R,{\cal P})+ \int_{\frac{xC-1}{\sqrt{Q}}}^{\infty} Dt \,C (t \sqrt{Q} + 1-\frac{xC}{2})
\, g(t;R,{\cal P}) \\
\label{eq.G1k2}
G_{C,2}(Q,R,x;{\cal P}) &= &\int_{\frac{-1}{\sqrt{Q}}}^{\infty} Dt \frac{C (t \sqrt{Q} + 1)^2}{1+2xC} \, g(t;R,{\cal P})
\end{eqnarray}

\noindent with

\begin{equation}
\label{eq.g}
g(t;R,{\cal P})= \int \frac{dy}{\sqrt{2 \pi (1-R^2)}} \,
\exp{\left(-\frac{\left(y \; {\rm sign}(P(y)) +t R \right)^2}{2(1-R^2)}\right)}.
\end{equation}

Deriving the free energy (\ref{eq.f}) with respect to $Q$, $R$
and $x$ gives three coupled equations for the order parameters.
These in turn, determine the properties of the SMC. The explicit expression of the saddle point equations for $k=1$ and $k=2$ is left to the Appendix, where we also derive
some general properties of the learning curves described in the
next sections.

The generalization error $\epsilon_g$, which is
the probability of misclassification of any pattern drawn with
probability (\ref{eq.probax}), is a geometric property that
depends only on $R$ and the rule to be learnt. In the case of
rules of type (\ref{eq.rule}), it is straightforward to obtain

\begin{equation}
\label{eq.eg}
\epsilon_g=\int Dt \, H \left(\frac{t R \; {\rm sign}({\cal P}(t))}{\sqrt{1-R^2}}\right)
\end{equation}

\noindent where $H(x)=\int_{x}^{\infty} Dt$. In the
particular case of a linearly separable rule ${\cal P}(z)=z$,
(\ref{eq.eg}) reduces to the usual expression
$\epsilon_g=\arccos(R)/\pi$.

The distribution of stabilities $\gamma_\mu \equiv h_\mu/\|{\bf w}\|$
of the training patterns, $\rho(\gamma)$, is given by

\begin{eqnarray}
\label{eq.rhoest}
\rho(\gamma)&=&\Theta(\gamma-\frac{1}{\sqrt Q}) \frac{e^{-\gamma^2/2}}{\sqrt{2 \pi}}g(-\gamma;R,{\cal P})+\Theta(\frac{1}{\sqrt Q} -\gamma)\frac{e^{-(\gamma-kxC)^2/2}}{\sqrt{2 \pi}}g(-\gamma+kxC;R,{\cal P}) \nonumber \\
&+&\delta(\gamma-\frac{1}{\sqrt Q}) (\int Dt \, (H\left(
\frac{t R \;{\rm sign}({\cal  P}(t))-\gamma}{\sqrt{1-R^2}}\right)-H\left(
\frac{t R \;{\rm sign}({\cal  P}(t))-\gamma+kxC/\sqrt{Q}}{\sqrt{1-R^2}}\right)).
\end{eqnarray}

The training error $\epsilon_t$ is the average fraction of
classification error on the training patterns. Integrating
(\ref{eq.rhoest}) over the negative stabilities we obtain

\begin{equation}
\label{eq.et}
\epsilon_t=\int Dt \, H\left(
\frac{t R \;{\rm sign}({\cal  P}(t))+kxC/\sqrt{Q}}{\sqrt{1-R^2}}\right).
\end{equation}

\noindent As expected, the training error is always strictly
smaller than the generalization error. Both converge to the
same limit for $\alpha \rightarrow \infty$.

\section{Learning curves}
\label{sec:learning}

In this section we present the learning curves, namely the
training error $\epsilon_t(\alpha)$ and the generalization
error $\epsilon_g(\alpha)$ of the SMCs for different teacher
rules. We include in the figures the learning curves of the
corresponding hard margin SVMs, or MSP, determined within the
hypothesis of replica symmetry. In the case of unrealizable
rules it is well known that the replica symmetry is broken
for $\alpha$ larger than $\alpha_{MSP}$, the fraction of
training patterns at which the hard margin $\kappa_{max}$,
positive for $\alpha < \alpha_{MSP}$, vanishes. The results
of computer simulations drawn on the same figures have been
obtained by solving numerically the dual problem~\cite{Martos}
using the Quadratic Optimizer for Pattern Recognition
program~\cite{AS}, that we adapted to the case without threshold
treated in this paper. The average has been taken over as many training sets as necessary (typically $\sim 500$ for small $\alpha$ and $\sim 50$ for big $\alpha$) to ensure that the error bars are smaller than the symbols. These simulations are in excellent
agreement with the theoretical predictions.

\subsection{The linear rule}
\label{sec:separable}

Introducing the expression ${\cal P}(z)=z$ corresponding
to a linearly separable teacher's rule in (\ref{eq.g}), we
obtain:

\begin{equation}
\label{eq.gsep}
g(t;R,{\cal P})= 2 \, H\left({\frac{Rt}{\sqrt{1-R^2}}}\right)
\end{equation}

The training and generalization errors, obtained after
solving the extremum equations for different values
of the hyperparameter $C$, are plotted against
$\alpha$ on Figures {\ref{fig.1} and {\ref{fig.2}
for $k=1$ and $k=2$ respectively. The generalization
error of the hard margin classifier,
solution of (\ref{eq.costprimMSP}) with conditions
(\ref{eq.condprimMSP}), and that of the optimal
bayesian generalizer~\cite{OpHa}, which are both
error-free solutions, are included on the figures
for comparison. Despite the fact that the task
is realizable by the student perceptron, the
training error for finite $C$ is finite. It goes through
a maximum and vanishes asymptotically in the limit
$\alpha \rightarrow \infty$. As expected, both for
$k=1$ and $k=2$ at any $\alpha$, $\epsilon_t$ is
larger the smaller the value of $C$,
which controls the relative importance of
the error term in the cost function
(\ref{eq.costprimSMC}). We can also see from the
figures that, given $C$, the machine with $k=2$
performs better than the one with $k=1$.
On increasing $C$, the learning curves approach
those of the MSP. In fact, by taking the limit
$C \rightarrow \infty$ in our saddle point equations
we get exactly the equations of the MSP for every
value of $\alpha$, independently of the power $k$.
This is not surprising, as in this limit the error
term dominates completely the soft margin cost function
(\ref{eq.costprimSMC}), which can only be minimized
if all the slack variables, and consequently the training
error, vanish. This is possible because the rule is
realizable. It is well known that the generalization error
of the MSP is larger than that of the bayesian generalizer
even asymptotically, as for $\alpha \rightarrow \infty$ both
algorithms have $\epsilon_g \sim a/\alpha$, but $a=0.5005$
in the case of the MSP~\cite{GoGr}, whereas $a=0.442$ for
the bayesian perceptron~\cite{OpHa}.

The obtained behaviour of the learning curves
at finite $C$ is reminiscent of that arising with
other learning algorithms having a hyperparameter.
In the inconsistent algorithms studied by Meir
and Fontanari~\cite{MeFo2}, patterns closer
to the hyperplane than a finite imposed distance
$\kappa > \kappa_{max}$ contribute to the cost,
linearly in the case of the perceptron algorithm and
quadratically in the case of the relaxation
one. In the algorithm Minimerror~\cite{GoGr}
the hyperparameter is equivalent to a learning
temperature. By training with these algorithms,
as well as with the SMC studied here, the generalization
error can be made smaller than that of the MSP by
choosing appropriate values for the hyperparameters,
at the price of learning with errors. The reason is
that, in contrast with the MSP, the bayesian solution
presents a finite fraction of training patterns at
any distance of the hyperplane~\cite{BuToGo}.
Thus, solutions with a small controlled fraction
of training errors may be closer to the
optimal bayesian hyperplane than the MSP, which has
no patterns at distances smaller than $\kappa_{max}$.

Unlike the generalization error of the inconsistent
learning algorithms, that vanishes asymptotically like
$\epsilon_g \sim 1/\sqrt{\alpha}$~\cite{MeFo2}, SMCs
with finite $C$ present a faster power law decay:

\begin{equation}
\label{eq.Egsep}
\epsilon_g \simeq \frac{\epsilon_0}{C^{\frac{1}{6}}}\, \frac{1}{\alpha^{\frac{2}{3}}},
\end{equation}

\noindent where the constant $\epsilon_0$ is
larger for $k=1$ than for $k=2$. In the limit
$C \rightarrow \infty$ eq. (\ref{eq.Egsep}) no
longer holds, and the well known decay
$\epsilon_g \approx \alpha^{-1}$ characteristic
of error-free trained perceptrons learning
realizable tasks is recovered.

Independently of the value of $C$, both the regularization term, proportional to $Q$, and the
slacks term diverge like $\sim \alpha^{2/3}$ for
$\alpha \rightarrow \infty$. In fact, this divergence
arises because we divided the free energy in (\ref{eq.elibre})
by $N$, instead of dividing by $N(1+\alpha)$, which
gives the energy per degree of freedom. In the large
$\alpha$ limit, this converges to $0$ as it should,
like $\alpha^{-1/3}$. The separable case is the only
one where the error term in the cost function presents
the same asymptotic behaviour as the regularization term.
In this limit, the soft margin $1/\sqrt{Q}$ vanishes
like $\alpha^{-1/3}$, in contrast with the hard margin
behaviour, $\kappa_{max} \approx \alpha^{-1}$~\cite{GoGr}.

\subsection{The shifted linear rule}
\label{sec:threshold}

Next we analyze the case of a linear teacher with a bias
$\delta>0$. The corresponding polynomial has a single root:
${\cal P}(z)=z-\delta$. This teacher separates linearly the
examples with a hyperplane at a distance $\delta$ from the
origin. As the student perceptron has no bias ($b=0$), zero
generalization error cannot be achieved: this rule is
unrealizable. The lowest value of $\epsilon_g$, obtained by taking
the asymptotic limit $R \rightarrow 1$ in (\ref{eq.eg}),
is $\epsilon_{g}^{\infty}=0.5-H\left(\delta\right)$.

The function $g$ defined by (\ref{eq.g}) is:

\begin{equation}
\label{eq.gthre}
g(t;R,{\cal P})= H\left({\frac{Rt+\delta}{\sqrt{1-R^2}}}\right)+H\left({\frac{Rt-\delta}{\sqrt{1-R^2}}}\right).
\end{equation}

\noindent Learning curves for different values of $C$
are represented as a function of $\alpha$ on Figure
\ref{fig.3}, for the particular value $\delta=0.3$.
The training error of the MSP is zero up to $\alpha_{MSP}$,
at which the maximal stability $\kappa_{max}$ vanishes.
$\alpha_{MSP}$ is a decreasing function of $\delta $. It
diverges at $\delta=0$, as the problem becomes separable,
and tends to the perceptron's capacity $\alpha_c=2$ in
the infinite $\delta$ limit. $\alpha_{MSP}$ cannot be
smaller than $\alpha_c$ since in the thermodynamic limit
any training set can be learned without errors for
$\alpha < \alpha_c$~\cite{Cover}. Within the replica
symmetry hypothesis, the MSP's training error $\epsilon_t$
displays a discontinuous transition at $\alpha=\alpha_{MSP}$.
For $\alpha > \alpha_{MSP}$, $\epsilon_t \equiv \epsilon_g$.
The generalization error does not present any singularity at
$\alpha=\alpha_{MSP}$. As already mentioned, $\kappa_{max}$
becomes negative for $\alpha>\alpha_{MSP}$, and the cost
function $E_{MSP}$ (\ref{eq.MSP}) is likely to present
several disconnected minima. Thus, the hypothesis of replica
symmetry used to draw the MSP's learning curves in Figure
\ref{fig.3} is most probably wrong.

If we take the limit $C \rightarrow \infty$ in our equations,
we get those corresponding to the MSP only for
$\alpha < \alpha_{MSP}$. At $\alpha_{MSP}$, the training error
of the SMC starts increasing and the generalization error curve
detaches down from that of the MSP, both through a second order
phase transition. The learning curves obtained in the limit
$C \rightarrow \infty$ are different for $k=1$ and $k=2$,
in contrast with the realizable rule considered before,
in which they converge to that of the MSP irrespective of the
value of $k$.

For finite values of $C$ the transition at $\alpha_{MSP}$
becomes a crossover both for $\epsilon_t$ and $\epsilon_g$,
at values of $\alpha<\alpha_{MSP}$ that decrease on decreasing
$C$. The training error for all $\alpha$ is larger than that
for infinite $C$, both for $k=1$ and $k=2$. The generalization
errors for different values of $C$ cross each other as a
function of $\alpha$. The envelope of the curves
$\epsilon_g(\alpha)$ corresponds to the lowest possible
value of $\epsilon_g$ reachable by the corresponding SMCs.
It depends on the exponent $k$. Notice that for large enough
values of $\alpha$ the replica symmetric approximation to
the MSP's generalization error is smaller than that of the
optimal soft margin solutions, and seems to provide a lower
bound to $\epsilon_g$ for the SMCs.

The convergence of the generalization error to its asymptotic limit,
for all values of $C$, is exponentially fast with $\alpha$:

\begin{equation}
\label{eq.Egthr}
\epsilon_g - \epsilon_g^{\infty} \simeq \exp(-\frac{\alpha}{a_k})
\end{equation}

\noindent The decay constant $a_k$ does not depend on $C$. A
stronger exponential drop of the generalization error, with
$\alpha^2$ in the exponent, has been found for SVMs learning
``easy" teacher rules. These not only are realizable, but present
a gap in the patterns distribution close to the discriminating
surface. In contrast, here the student's hyperplane is surrounded
by unlearnable patterns. The student cannot get rid of the errors
by decreasing the soft margin, like with the linear rule. On
increasing $\alpha$, ${Q}$ converges to a constant that depends on
$k$ and $\delta $ while the error term in (\ref{eq.costprimSMC})
increases with $\alpha$. For large enough $\alpha$, the cost
function is mainly dominated by the error term, and then $C$ only
plays the role of an irrelevant multiplicative constant. This is
why the convergence rate to the asymptotic value of the
generalization error does no depend on $C$.

Similar results are obtained for $k=2$, as is shown
on Figure \ref{fig.4}.

\subsection{Sandwich Rule}
\label{sec:sandwich}

Consider now rules of the form ${\cal P}(z)=z(z-\delta)$,
where the polynomial defining the teacher's output has two
roots. The corresponding discriminating surfaces are two
parallel hyperplanes, one containing the origin and the
other at a distance $\delta/\sqrt N$ of it. The patterns that lying
between the hyperplanes belong to class $+1$, the others to
class $-1$. Thus, not only these are unrealizable rules, but
the classification errors will necessarily correspond to
patterns at large distance of the student's hyperplane.

As with all the unrealizable rules, the training error of the
MSP within the replica symmetric approximation presents a
discontinuity at $\alpha_{MSP}$ where $\kappa_{max}$ vanishes.
Here $\alpha_{MSP}$ is an increasing function of $\delta$,
starting at $\alpha_{MSP}=2$ for $\delta =0$, which corresponds
to the most difficult learning task and diverging for
$\delta \rightarrow \infty$. The generalization error starts
decreasing at small $\alpha$, reaches a minimum beyond
$\alpha_{MSP}$ and then starts to increase, and tends
asymptotically to $\epsilon_g=1/2$ for $\alpha \rightarrow
\infty$. Notice however that for $\alpha > \alpha_{MSP}$
the replica symmetry is most probably broken.

The properties of the SMC are obtained by replacing

\begin{equation}
\label{eq.gsan}
g(t;R,{\cal P})= 2H\left({\frac{Rt}{\sqrt{1-R^2}}}\right)+H\left({\frac{\delta -Rt}{\sqrt{1-R^2}}}\right)-H\left({\frac{Rt+\delta}{\sqrt{1-R^2}}}\right),
\end{equation}

\noindent in the saddle point equations (\ref{eq.1}-\ref{eq.3})
of the Appendix.

The learning curves for different values of the hyperparameter
$C$, corresponding to a width $\delta=2$, are represented on
Figures \ref{fig.5} and \ref{fig.6} for $k=1$ and $k=2$
respectively. Given $C$, for large enough $\alpha$, the training
error curves $\epsilon_t(\alpha)$ for $k=1$ are below those for
$k=2$. This is so because the unavoidable errors, which are very
far from the hyperplane, are more heavily penalized if $k=2$.
Thus, the SMC tries to learn these examples even if this increases
the overall number of errors. As a result, learnable patterns
close to the hyperplane, that have small slacks, are incorrectly
classified. This can be checked up by taking a look at the
distribution of stabilities, Figure \ref{fig.7}.

Like with the previous shifted linear rule, the norm of
the student's weight vector $Q$ tends to a constant value
and therefore, the error term dominates the cost function
in the asymptotic limit $\alpha \rightarrow \infty$. However,
instead of the exponential convergence, the generalization
error decays asymptotically to $\epsilon_g^{\infty} =
H\left(\delta\right)$ like $\alpha^{-\frac{1}{2}}$. The
reason of this difference is discussed in section
\ref{sec:Discussion}.

\subsection{The Reversed Wedge}
\label{sec:Reversed}

Teachers defined by third order polynomials like ${\cal P}(z)= z
(z-\delta)(z+\delta)$ with $\delta > 0$, correspond to the so
called Reversed Wedge~\cite{WR} rules. Patterns ${\bf x}_\mu$ with
${{\bf w}_0} \cdot{{\bf x}_\mu} \in  (-\infty,-\delta) \cup
(0,\delta)$ belong to class $-1$, those outside this subspace to
class $+1$. The generalization properties of a perceptron learning
a reverse wedge teacher have been addressed in~\cite{WR}, and
within the on-line paradigm, using Hebb's learning rule
in~\cite{INK}.

The behaviour of the replica symmetric approximation to the MSP is
as described in the previous section, but here $\alpha_{MSP}$
diverges both in the limits of vanishing and infinite wedge width
$\delta$, for which the problem becomes separable, and has a
minimum at $\delta_c=\sqrt{2 \ln 2}$~\cite{WR}. At this value of
$\delta$ the patterns stability distribution along the teacher's
weight $\bf w_0$ has zero mean. Correspondingly, learning becomes
impossible for the MSP, as is discussed later. Thus, for
$\delta_c$, $R=0$ for every value of $\alpha$, and
$\alpha_{MSP}=2$ is equal to the perceptron's capacity.

The properties of the SMCs are deduced
after insertion of

\begin{equation}
\label{eq.grw}
g(t;R,{\cal P})= 2 H\left({\frac{Rt -\delta }{\sqrt{1-R^2}}}\right)+H\left({\frac{Rt+\delta}{\sqrt{1-R^2}}}\right)- H\left({\frac{Rt}{\sqrt{1-R^2}}}\right)
\end{equation}

\noindent into the saddle point equations.

In contrast with the problems considered before,
the generalization error of a perceptron learning the
reversed wedge rule is a monotonic function of $R$
{\it only if} $\delta > \delta_c$~\cite{INK}. For
$0 < \delta < \delta_c$, $\epsilon_g(R)$ presents
a relative minimum at $R_{min}>0$ and a corresponding
maximum at $-R_{min}$. The relative minimum is the global
one only for $0<\delta<\delta_* \equiv 0.570185$.
At $\delta_*$ the global minimum jumps to $R=-1$, and
for $\delta_*<\delta<\delta_c$ the generalization error
takes its smallest value at $R=-1$. At $\delta=\delta_c$ the
relative extrema collapse at the inflexion point $R_{min}=0$,
and for larger values of $\delta$ the generalization
error becomes a monotonic {\it increasing} function of $R$.
This behaviour is represented on Figure \ref{fig.8}.

For the values of $k$ investigated, $R$ has two distinct
behaviors as a function of $\alpha$, depending on the
wedge's width $\delta$. If $\delta<\delta_c$, the teacher's
average stability is positive, and $R(\alpha)$
is a monotonic continuous function growing from $0$ to its
asymptotic value $+1$. In this range of small wedges, the
soft margin learning algorithm does not converge to the
minimal value of the generalization error in the limit of
infinite $\alpha$, as is the case in the other tasks
considered before. In fact it ``overshoots'', in the sense
that $R(\alpha)$ continues to grow beyond the value that
optimizes the generalization performance. Correspondingly,
$\epsilon_g(\alpha)$ goes through a minimum at finite
$\alpha$ but, as $R$ increases with $\alpha$, it converges
to a larger value, $\epsilon_g^\infty \equiv \epsilon_g(R=1)$.
The learning curves of Figure \ref{fig.9} are an example
of this behaviour. Notice that for $\delta_*<\delta<\delta_c$
this value of $\epsilon_g^\infty$ corresponds
the {\it largest} value of the student's generalization
error. Moreover, for $0.67449<\delta<\delta_c$ the asymptotic
behaviour is even worse than a random guess, because
$\epsilon_g(R=1)>0.5$.

At $\delta=\delta_c$ there is an abrupt change of the learning
behaviour, as beyond this wedge's width the average teacher's
stability is negative, and $R$ becomes a decreasing function of
$\alpha$. Correspondingly, the soft margin solution converges to
the optimal generalizer in the limit $\alpha \rightarrow \infty$.
This corresponds to $R=-1$, because for large $\delta$, most of
the patterns lie in inside the reversed wedge, so that the
student's weight vector tends to orient {\it antiparallel} with
the teacher's vector ${\mathbf w}_0$, in order to classify
correctly most of the examples. Learning curves for $\delta=2 >
\delta_c$ obtained with exponent $k=1$ for the slacks exponent in
the cost function are represented on Figure \ref{fig.10}.

As for the sandwich rule, the generalization error
decays as $\alpha^{-\frac{1}{2}}$ to the corresponding
asymptotic values, $\epsilon_g^{\infty}=1-2H(\delta)$ for
$R \rightarrow 1$, and $\epsilon_g^{\infty}=2H(\delta)$ for
$R \rightarrow -1$. The same asymptotic behaviors for
$\epsilon_g$ and $R$, but with different prefactors, were
obtained by Inoue et al.~\cite{INK} for the online Hebbian
learning scenario.

The asymptotic value of $Q$ tends to zero as $\delta$ tends
to $\delta_c$. In the two limiting cases $\delta \rightarrow
\infty$ and $\delta \rightarrow 0$, the task becomes
linearly separable and correspondingly $Q \rightarrow
\infty$.

For the particular case of $\delta = \delta_c$, the only
solution of the saddle point equations is $R=0$ for every
value of $\alpha$. This ``no learning" regime is discussed
in section \ref{sec:Discussion}.

\section{Optimization of the hyperparameter}
\label{sec:Copt}

The figures of the preceding section show that the
behavior of the generalization error of the SMC
is not monotonic with $C$. It can be seen that there
is an optimal value $C_{opt}(\alpha)$ that allows
to obtain the minimum generalization error for each
$\alpha$. Obviously, $C_{opt}$ cannot be calculated
using the training examples alone, so that in the
applications it can only be estimated. Several methods
for doing this have been proposed recently~\cite{Sollich,Seeg}.
Here we determine the statistical properties of the
optimal SMC, thus providing reference curves against
which results obtained using the different estimators
may be tested.

As $\epsilon_g$ depends implicitly on $C$ through $R$,
in the cases where $\epsilon_g$ is a monotonic function
of $R$, its minimum is obtained by looking for the
extremum of $R$ with respect to $C$, at fixed $\alpha$,
$C_{opt}(\alpha)$. To this end, the three saddle point
equations (\ref{eq.1}-\ref{eq.3}) of the Appendix, together
with their derivatives with respect to $C$, constitute
a system of 6 coupled equations for the variables $Q$,
$R$, $x' \equiv xC$, ${\partial Q}/{\partial C}$,
${\partial R}/{\partial C}$ and ${\partial x'}/{\partial C}$.
Setting the extremum condition $\partial R/\partial C=0$,
the equations obtained by derivation of (\ref{eq.2}) and
(\ref{eq.3}) form a homogeneous system for
$\partial Q/\partial C$ and $\partial x'/\partial C$.
The only nontrivial solution is obtained by setting the
determinant of this system to zero, which gives

\begin{equation}
\label{eq.coptimo}
\frac{\partial^2 f}{\partial Q \partial R} \cdot \frac{\partial^2 f}{\partial
x^2} - \frac{\partial^2 f}{\partial R \partial x} \cdot \frac{\partial^2
f}{\partial Q \partial x} = 0.
\end{equation}

\noindent where $f$ stands for the free energy
(\ref{eq.f}). Solving the system given by equation
(\ref{eq.coptimo}) together with the three original
saddle point equations for $Q$, $R$,
$x'$ and $C$, we get $C_{opt}$ in the cases where
$\epsilon_g$ is a monotonic function of $R$.

In the other cases, as happens with the
Reverse Wedge rule, determining $C_{opt}$
is less straightforward because the minimum
of $\epsilon_g$ may be reached for a value
$R^*$ (different from $\pm 1$) such that
$\partial \epsilon_g / \partial R (R^*)=0$,
with $\partial R/\partial C \neq 0$. In that
case, $C_{opt}$ is the one that gives
$R(C_{opt})=R^*$, and has to be determined
numerically.

The optimal generalization curves for the different rules
considered in this paper are represented on the figures of the
preceding section. Notice that for $\alpha<\alpha_{MSP}$, the MSP
is not optimal for any value of $\alpha$, as it is obtained in the
limit $C \rightarrow \infty$. In the case of the realizable linear
separation, the optimal generalization error of the SMC vanishes
asymptotically as $0.488\alpha^{-1}$ for $k=1$, and as
$0.449\alpha^{-1}$ for $k=2$. The latter is very close to that of
the bayesian perceptron, $0.442\alpha^{-1}$, but the curves are
also very close for finite values of $\alpha$, as can be seen on
figure \ref{fig.2}. Notice that the asymptotic decay of
$\epsilon_g$ for the SMC is faster than that of the MSP, even for
$k=1$. This is an interesting result, as it shows that, even when
a hard margin solution exists, learning with a soft margin machine
allows to obtain better classifiers.

For the non separable cases, even if $C_{opt}$ allows
to obtain the best performances at finite $\alpha$,
since the asymptotic behavior of $\epsilon_g$ is
independent of $C$, all the learning curves, including
the optimal one, tend to a value that only depends on
the rule and on $k$, as shown in the corresponding
sections.

The evolution of $C_{opt}$ with $\alpha$ can be
seen on Figures \ref{fig.11} to \ref{fig.12}.
The behaviour of the curves is qualitatively similar
for the shifted linear rule and the reversed wedge with
small $\delta$ on one hand, and for the sandwich
rule and the reversed wedge with large $\delta$ on
the other. The divergences of $C_{opt}$ are related
to the presence of errors with unbounded slack values.
For $\alpha$ beyond the divergence, $C_{opt}=\infty$.

\section{Discussion}
\label{sec:Discussion}

In the preceding sections we presented the learning curves
of a SMC learning a variety of rules, characterized by an
anisotropy axis parallel to the teacher's vector ${\bf w}_0$.
Some of the obtained results, and in particular the asymptotic
behaviour in the $\alpha \rightarrow \infty$ limit, can
be generalized to other teacher rules (Proofs are detailed
in the Appendix). As shown by Reimann and Van den
Broeck~\cite{REVdB}, it is useful to characterize
the teacher rules by the average patterns' stability of a
perceptron aligned with the teacher's vector,

\begin{equation}
\label{eq.avstab}
\langle \gamma \rangle = \int d\gamma \, \rho(\gamma) =
\int Dz \,z \,{\rm sign}({\cal P}(z)),
\end{equation}

\noindent where the second equality in (\ref{eq.avstab})
stems from our assumption (\ref{eq.probax}) that the
patterns' distribution is a gaussian.

In the Appendix we show that in the limit $\alpha
\rightarrow \infty$, both for $k=1$ and $k=2$, $R$
converges asymptotically either to $1$ or to $-1$,
that is, the student perceptron gets either completely
aligned or completely anti-aligned with teacher's vector.
Furthermore, for non separable rules, $1-R^2 \sim 1/\alpha$.
In this limit of $R \rightarrow \pm 1$ we find $\epsilon_g
\rightarrow \epsilon_g^{\infty}=\int Dz \theta(\mp z \, {\cal P}(z))$,
irrespective of the teacher's rule. The convergence law to this
asymptotic value depends on whether the polynomial ${\cal P}(z)$
defining the rule in (\ref{eq.rule}) has or not a root $z_i=0$.
If $0$ is {\it not} a root of ${\cal P}(z)$, ${\cal P}(0) \neq 0$
and $\epsilon_g -\epsilon_g^{\infty} \sim
\exp(-{\rm \varepsilon }/(1-R^2))$ with $\varepsilon$ a constant, whereas if $0$ is a root, then
the decay follows the law $\epsilon_g -\epsilon_g^{\infty}
\sim \sqrt{1-R^2}$.

Thus, for the unrealizable rules that have $0$ as one of the
roots of ${\cal P}$, the generalization error decays to the
asymptotic value as $\epsilon_g-\epsilon_g^{\infty} \sim
\alpha^{-\frac{1}{2}}$. A similar result has been obtained
by Amari et al.~\cite{AFS} within the annealed approximation
for the case of a deterministic machine learning a noisy
teacher, and by other authors for hebbian learning of
unrealizable tasks~\cite{INK,SST}. The same power law has
been obtained by Meir and Fontanari~\cite{MeFo2} for
a realizable problem learned with inconsistent algorithms,
within the approximation of replica symmetry, which is probably
not valid for large values of $\alpha$. Indeed, the
soft margin algorithm with finite $C$ is also inconsistent
when the rule is the linear separation considered in section
\ref{sec:separable}, and in that case we obtain a different
power law decay.

In the case of a linearly separable rule, the SMC with
$C_{opt}$ has $\epsilon_g \approx 1/\alpha$, like the MSP,
which corresponds to $C=\infty$. However, at fixed finite
values of $C$ the decay is slower, like $\sim 1/\alpha^{2/3}$.
The same exponent has been obtained for a perceptron learning
a separable rule using noisy examples with one step of replica
symmetry breaking~\cite{UeKa}. Within the replica symmetric
approximation to the same problem the exponent is $1/2$ instead
of $2/3$~\cite{GyTi}.

In the cases where $0$ is not a root of ${\cal P}(z)$, like for
the shifted linear rule, the decay is exponential,
$\epsilon_g-\epsilon_g^{\infty} \sim \exp(-\varepsilon \cdot
\alpha)$. A similar behaviour was found in \cite{SST} for a
perceptron with linear output and binary weights, trying to learn
examples given by a teaches with the same structure, where the
generalization error vanishes exponentially.

The presence or the absence of a root $z_i=0$ induces
different asymptotic behaviors because if $0$
is a root, then a student perceptron aligned with the teacher
has $|R|=1$ and can perfectly separate the patterns closest
to the hyperplane. In that case, any small misalignement
modifies the classification induced by the student, thus
strongly modifying the error term in the cost function.
On the other hand, if $0$ is not a root, the student's
hyperplane is immersed in a sea of patterns of the same
class. Small tilts of the hyperplane do not change significantly
the classification nor the slacks term in the cost.

It is interesting to notice that the figures of the
learning curves as well as those of $C_{opt}$ show
an analogy between the behaviour for the SMCs
with bounded slacks, like in the case of the shifted linear
rule and that of the reversed wedge when $\delta <\delta_c$,
and between those with unbounded slacks, as is the case with
the sandwich rule and the reversed wedge when $\delta >\delta_c$.
For this last type of rules, $C_{opt}$ diverges beyond some finite
$\alpha$.

Consider now the small $\alpha$ limit. As
shown in the Appendix, $R \sim \langle \gamma \rangle \,
\sqrt{\alpha}$ and so, $\epsilon_g \sim 1/2-\langle
\gamma \rangle^2 \sqrt{\alpha}$. Thus, irrespective of
the rule considered, when the fraction of training
examples is small, the SMC generalizes better than
by random guessing. This is not necessarily the case
for larger values of $\alpha$.

If we put $R=0$ in the equations, and solve
for $\alpha$, the only possible solution when
$\langle \gamma \rangle \neq 0$, is $\alpha =0$.
Thus, $R \neq 0$ for all $\alpha$, and
has the sign of $\langle \gamma \rangle$ unless it has discontinuous changes of sign. Notice that, given the asymptotic behaviours just mentioned, if $R$ is discontinuous it can only have an even number of changes of sign. A similar
result has already been obtained in a broader
frame~\cite{REVdB}. From the behaviour of $R$ in the small $\alpha$ limit,
it can be seen that the problem gets very
difficult to learn for rules with $\langle \gamma \rangle$
close to $0$. In fact, in the very special case of
$\langle \gamma \rangle=0$, $R=0$ is a solution of
the saddle point equations for every value of $\alpha$.
If this is the only solution, the machine cannot learn
at all, as is the case for the reverse wedge rule when
$\delta=\delta_c$. This behaviour is similar
to the one of retarded learning, found in
problems of unsupervised learning with
quadratic cost functions~\cite{REVdB}. In that case,
it has been shown that learning is still possible,
provided that the cost function is capable to
extract the information about the anisotropy
of the distribution of stabilities, contained
in its higher order moments~\cite{BuGo}. Notice
that this is not the case for the cost functions
for the SMCs considered in this paper.

\section{Conclusion}

The properties of the recently proposed Support
Vector Machines have been studied theoretically
in two situations of interest, namely for the cases
where the student has either the same structure as the
teacher, or it is more complex than it. In both
situations the rule to be learned is realizable, and
interesting properties of hard margin SVMs, like the
existence of hierarchical generalization, could be
analyzed within the replica symmetry hypothesis~\cite{DOS}.

In the present paper we addressed the situation where the task is
more complex than the learning machine. In this case the cost
function for the SVMs is modified. It allows to obtain a Soft
Margin Classifier that results from a trade-off, controlled by a
single parameter $C$, between increasing the margin and minimizing
the number of training errors. As the cost function is quadratic
and the domain of solutions is convex, we obtain the typical
learning curves for a variety of unrealizable tasks using the
replica symmetry hypothesis. We considered problems characterized
by a single symmetry-breaking direction ${\bf w_0}$, along which
the patterns have alternating positive or negative class label. We
have shown that the convergence of the corresponding learning
curves to the asymptotic value follows either a power law or an
exponential, depending on the position of the singularities of the
teacher's rule.

Even if the student is well adapted to the
task's complexity, the SMC may generalize
better than the error-free hard margin SVM,
provided the hyperparameter $C$ in the cost
function is correctly tuned. It can even attain almost Bayesian performance.

We showed that the prefactors of the different asymptotic
behaviours are proportional to the average
stability of the teachers rule,
$\langle \gamma \rangle$. When this vanishes,
the SMC with cost function (\ref{eq.costprimSMC})
cannot learn, and the overlap between the
student and the teacher directions is $R=0$.
We considered two exponents for the error term
in the cost function, $k=1$ and $k=2$.
It would be interesting to study the properties
of SMCs trained using exponents $k>2$ in the
cost function, as we expect that these should
detect the difference of the odd moments
of the patterns distribution in the
directions parallel and orthogonal to
${\bf w}_0$.

Another interesting question is whether the
hierarchical learning of hard margin SVMs exists
also with SMCs. To tackle this question, pattern
distributions with two different anisotropies have
to be considered.

\section{Acknowledgments}
\label{sec:Acknowledgments}

We thank Alex Smola for providing us with the Quadratic Optimizer for Pattern Recognition program~\cite{AS}. SR-G acknowledges economic support from the EU-research contract ARG/B7-3011/94/97. It is a pleasure to acknowledge support from the Zentrum f\"ur interdisziplin\"are Forschung in Bielefeld, where this work was finished in the frame of the Research Group "The Sciences of Complexity: From Mathematics to Technology to a Sustainable World".

\section{Appendix}
\label{sec:Appendix}

The saddle point equations for the cases $k=1$ and $k=2$ are:

\begin{eqnarray}
\label{eq.1}
1-R^2-x  &=& \alpha \cdot I_1(xC, \sqrt{Q},R;k),
\\
\label{eq.2}
R &=& -\alpha \cdot I_2(xC, \sqrt{Q},R;k),
\\
\label{eq.3}
1-R^2 &=& \alpha  \cdot I_3(xC, \sqrt{Q},R;k),  ,
\end{eqnarray}

\noindent with, for the case $k=1$

\begin{eqnarray}
\label{eq.I1k1}
I_1(xC,Q,R;1) &=& \int_{\frac{-1}{\sqrt{Q}}}^{\frac{xC-1}{\sqrt{Q}}} Dt \, t\, (t+\frac{1}{\sqrt{Q}})g(R,t,P)+
\int_{\frac{xC-1}{\sqrt{Q}}}^{\infty } Dt \, \frac{txC}{\sqrt{Q}} g(R,t,P),
\\
\label{eq.I2k1}
I_2(xC,Q,R;1) &=& \int_{\frac{-1}{\sqrt{Q}}}^{\frac{xC-1}{\sqrt{Q}}} Dt  \frac{1}{2} \, (t+\frac{1}{\sqrt{Q}})^2 \frac {\partial g(R,t,P)}{\partial R}+
\int_{\frac{xC-1}{\sqrt{Q}}}^{\infty } Dt \frac{xC}{\sqrt{Q}} (t+ \frac{2-xC}{2 \sqrt{Q}}) \frac{\partial g(R,t,P)}{\partial R},
\\
\label{eq.I3k1}
I_3(xC,Q,R;1) &=& \int_{\frac{-1}{\sqrt{Q}}}^{\frac{xC-1}{\sqrt{Q}}} Dt \, (t+\frac{1}{\sqrt{Q}})^2 g(R,t,P)+
\int_{\frac{xC-1}{\sqrt{Q}}}^{\infty } Dt \frac{(xC)^2}{Q} g(R,t,P)
\end{eqnarray}

\noindent and, for the case $k=2$,

\begin{eqnarray}
\label{eq.I1k2}
I_1(xC,Q,R;2) &=& \int_{\frac{-1}{\sqrt{Q}}}^{\infty}Dt \frac{2xCt}{1+2xC} \, (t+\frac{1}{\sqrt{Q}}) \,g(R,t,P)
\\
\label{eq.I2k2}
I_2(xC,Q,R;2) &=& \int_{\frac{-1}{\sqrt{Q}}}^{\infty } Dt  \frac{xC}{1+2xC} \, (t+\frac{1}{\sqrt{Q}})^2 \frac {\partial g(R,t,P)}{\partial R}
\\
\label{eq.I3k2}
I_3(xC,Q,R;2) &=& \int_{\frac{-1}{\sqrt{Q}}}^{\infty } Dt \, \frac{(2xC)^2}{(1+2xC)^2} \, (t+\frac{1}{\sqrt{Q}})^2 g(R,t,P)
\end{eqnarray}

From (\ref{eq.I3k2}) it can be seen that, for $k=2$, $x$ must vanish in the infinite $\alpha$ limit in order to make $I3$ vanish. Notice that the function $g(R,T,P)$ is always nonnegative (\ref{eq.g}). For the case $k=1$ the analysis of (\ref{eq.I3k1}) shows that $x$ must either vanish or tend to a positive constant with $q$ tending to infinity. This last case can be ruled out by noticing that it is inconsistent with the vanishing of $I2$ (notice that (\ref{eq.I2k1}),as well as (\ref{eq.I2k2}) can be solved analitically).

To show that $R$ can only tend to $1$ or $-1$ in the infinite $\alpha$ limit, it is useful to rewrite I1 and I2, which in the case $k=1$ are

\begin{eqnarray}
\label{eq.I1k1b}
I_1(xC,Q,R;1) &=& \int_{\frac{-1}{\sqrt{Q}}}^{\frac{xC-1}{\sqrt{Q}}} Dt \, g(R,t,P) + R \, I_2(xC,Q,R;2)
\\
\label{eq.I2k1b}
I_2(xC,Q,R;1) &=& \sum_{i=1}^N \tau(x_i^+) \frac{e^{-x_i^2}}{\sqrt{2 \pi}} \{\int_{\frac{{\frac{-1}{\sqrt{Q}}-x_i R}}{\sqrt{1-R^2}}}^{\frac{{\frac{xC-1}{\sqrt{Q}}-x_i R}}{\sqrt{1-R^2}}} Dt \, (t \sqrt{1-R^2}+\frac{1}{\sqrt Q}+ x_i R)
\nonumber \\
&+& \frac{xC}{\sqrt Q} \int_{\frac{{\frac{xC-1}{\sqrt{Q}}-x_i R}}{\sqrt{1-R^2}}}^{\infty} Dt+ (x_i \leftrightarrow -x_i) \}
\end{eqnarray}

\noindent and for $k=2$,

\begin{eqnarray}
\label{eq.I1k2b}
I_1(xC,Q,R;2) &=& \frac{2xC}{1+2xC} {\int_{\frac{-1}{\sqrt Q}}^{\infty } Dt \, g(R,t,P) + R \, I_2(xC,Q,R;2)}
\\
\label{eq.I2k2b}
I_2(xC,Q,R;2) &=& \frac{-2xC}{1+2xC} \sum_{i=1}^N \tau(x_i^+) \frac{e^{-x_i^2}}{\sqrt{2 \pi}} \{\int_{\frac{{\frac{-1}{\sqrt{Q}}-x_i R}}{\sqrt{1-R^2}}}^{\infty } Dt \, (t \sqrt{1-R^2} \nonumber \\
&+& \frac{1}{\sqrt Q}+ x_i R) + (x_i \leftrightarrow -x_i) \}.
\end{eqnarray}

Let us suppose that $R$ tends to a constant different from $1$ and $-1$ as $\alpha$ tends to infinity. It can be seen that in that case $I1$, $I2$ and $I3$ must vanish at the {\it same} rate. If we consider teachers with at least one positive root , i. e. {\it unrealizable} teachers, it can be seen that the integral in I3 (the second one for the case k=1) never vanishes. Thus, $I3$ must vanish as $(x/\sqrt Q)^2$ for k=1 and as $x^2$ for k=2, if $Q$ tends to a constant or to infinity. But equations (\ref{eq.I1k1b}) and (\ref{eq.I1k2b}) show that $I1$ and $I2$ cannot vanish at the same rate as $I3$ because the first term on the right handside vanishes as $x/\sqrt Q$  for $k=1$ and as $x$ for $k=2$. If $Q$ tends to $0$ then I3 must vanish as $(x/\sqrt Q)^2$ for both cases. But then $I2$ cannot vanish at the same rate, because equations (\ref{eq.I2k1b}) and (\ref{eq.I2k2b}) show that $I2$ must vanish as $xC \langle \gamma \rangle/\sqrt Q$, unless $\langle \gamma \rangle =0$ (this case will be analyzed below). Therefore, $R$ tends either to $1$ or to $-1$ for all teachers with $\langle \gamma \rangle \not =0$.

By putting $R=0$ in the equations one can easily (notice that $g(0,t,P) \equiv 1$) see that if $ \langle \gamma \rangle \not =0$, it can only be a solution for $\alpha=0$. On the other hand, for $\langle \gamma \rangle =0$, $R=0$ is a solution for {\it every} value of $\alpha$, i. e. learning is impossible for this kind of teacher.

It is also possible to find the condition that makes $R$ go to each one of its limiting values (1 or -1). From what has been said before regarding $I3$ it can be seen that it vanishes as $x^2$, and so, $1-R^2 \sim \alpha x^2$. Using this, and equation (\ref{eq.1}) it is evident that $I1$ must vanish faster than $x$. But, in the infinite $\alpha$ limit, $I1$ is written, to first order,

\begin{eqnarray}
\label{eq.I1k1as} I_1(xC,Q,R;1) &\sim& \frac{-x}{\sqrt Q} \{
sign(R) \langle \gamma \rangle + \int_{-\infty }^{\frac{-1}{\sqrt
Q}} Dt \, t \, g(\pm 1,t,P) \}
\\
\label{eq.I1k2as}
I_1(xC,Q,R;2) &\sim& -x \{ sign(R) \frac {\langle \gamma \rangle}{\sqrt Q}  - \int_{\frac{-1}{\sqrt Q}}^{\infty}  Dt g(\pm 1,t,P) \} \nonumber \\
&-& sign(R) \sum_{i/ {|x_i|>\frac{1}{\sqrt Q}}}  \tau(x_i^+)
(|x_i|- \frac{1}{\sqrt Q}) \frac{e^{-x_i^2}}{\sqrt{2\pi}} \}
\end{eqnarray}

Thus, the term within brackets must vanish. For (\ref{eq.I1k1as}) it is evident that this can only happen if $R \longrightarrow sign(\langle \gamma \rangle)$. The same can be shown for (\ref{eq.I1k2as}), with a bit of algebra. The asymptotic value of $Q$ can be obtained by imposing the vanishing of the above mentioned terms.

To see the rate of decay of $1-R^2$, notice that, from (\ref{eq.3}) and from the fact (shown above) that $I3 \sim x^2$, one gets that $1-R^2 \sim \alpha x^2$. But, using the fact that $I1$ must decay faster than $x$, equations (\ref{eq.I2k1b}) and (\ref{eq.I2k2b}) impose that $I2 \sim x$. This, together with (\ref{eq.2}), gives that $x \sim 1/\alpha$. Therefore, $1-R^2 \sim 1/\alpha$.

\pagebreak

\begin{figure}
\centerline{\psfig{figure=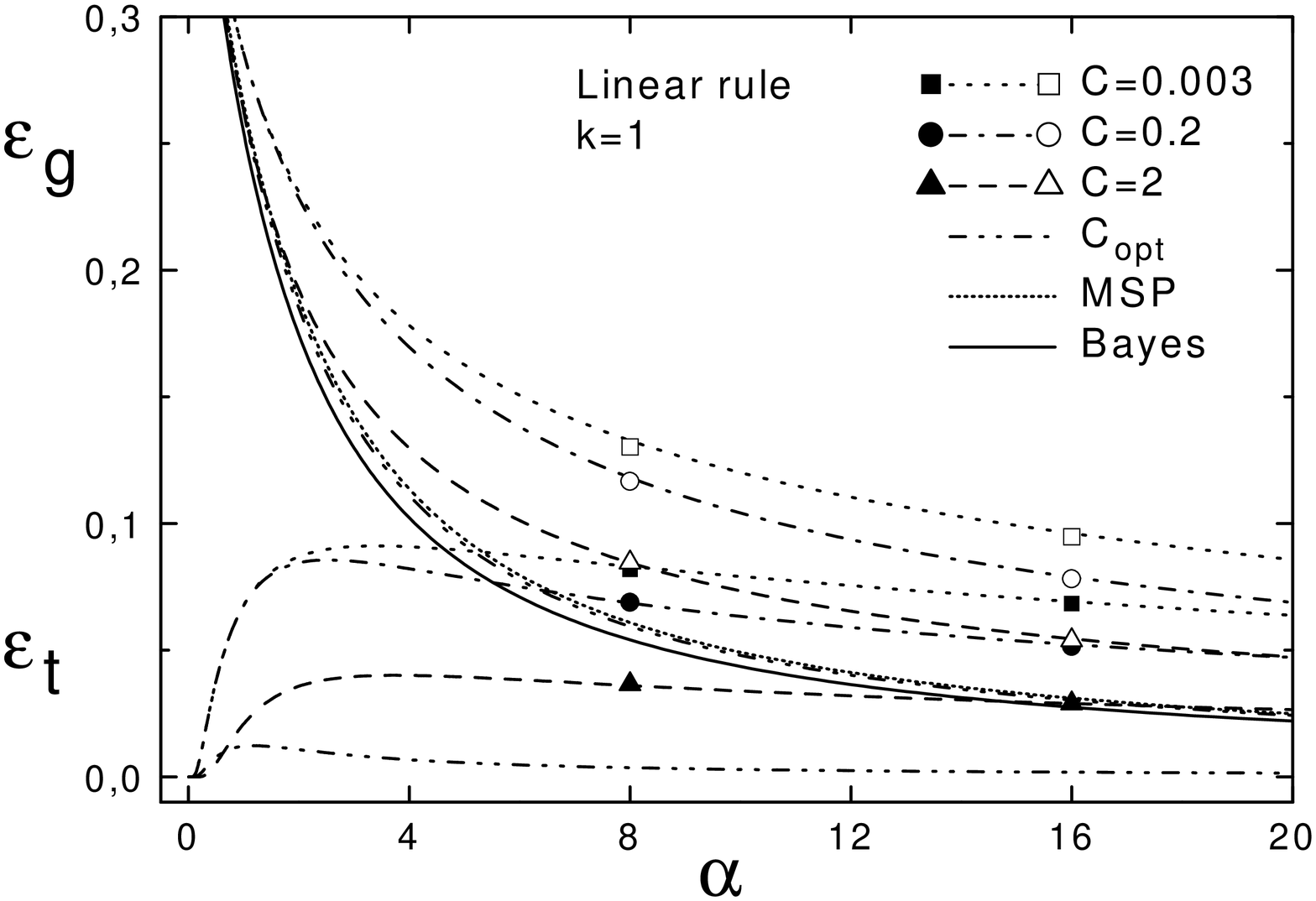,height=7cm}}
\caption{Linearly separable rule. SMC's learning
curves ($\epsilon_t$ below, $\epsilon_g$ above)
corresponding to an exponent $k=1$ in the cost
function, for different values of the hyperparameter
$C$. The generalization errors of the MSP and the
optimal (bayesian) generalizer, are included for
comparison. The learning curves of the optimal SMC,
discussed in section \ref{sec:Copt}, are also
represented. Symbols, $\epsilon_t$ in black,
$\epsilon_g$ in white, correspond to results of computer
simulations with $N=100$. Error bars are smaller than the
symbols.}
\label{fig.1}
\end{figure}

\pagebreak

\begin{figure}
\centerline{\psfig{figure=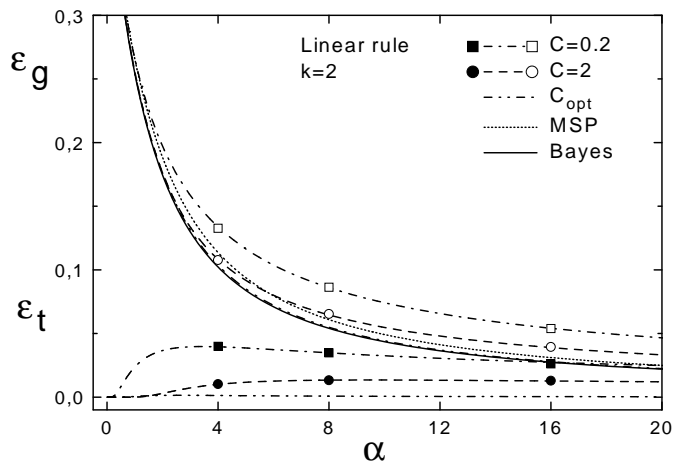,height=7cm}}
\caption{Linearly separable rule. Same as the preceding
figure, with an exponent $k=2$ in the cost
function.}
\label{fig.2}
\end{figure}

\pagebreak

\begin{figure}
\centerline{\psfig{figure=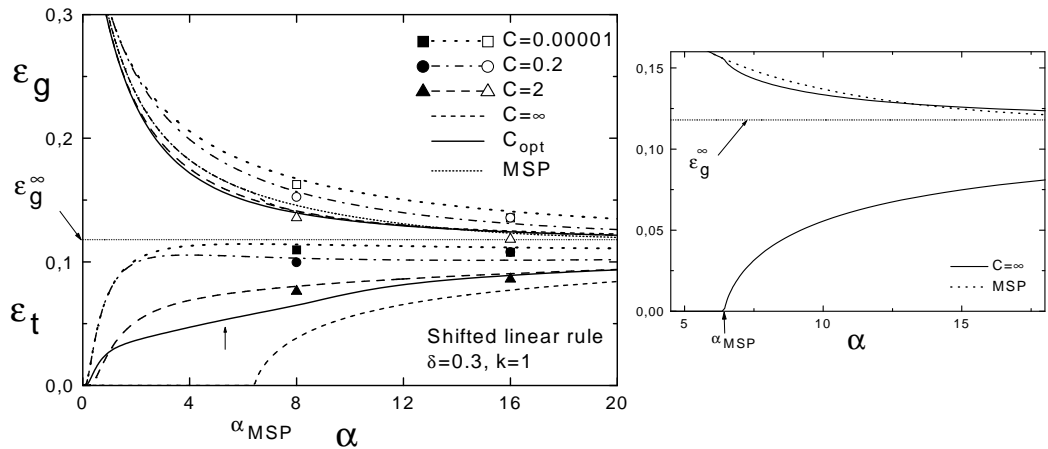,height=7 cm}}
\caption{Shifted linear rule. SMC's learning
curves corresponding to an exponent $k=1$ in the cost
function, for different values of the hyperparameter
$C$. Symbols correspond to results of computer
simulations with $N=50$. Error bars are smaller than the
symbols. The figure in the right shows the difference
between the MSP within the replica symmetry approximation
and the SMC learning with $C=\infty$. Asymptotically,
$\epsilon_g^\infty=0.1179$.}
\label{fig.3}
\end{figure}

\pagebreak

\begin{figure}
\centerline{\psfig{figure=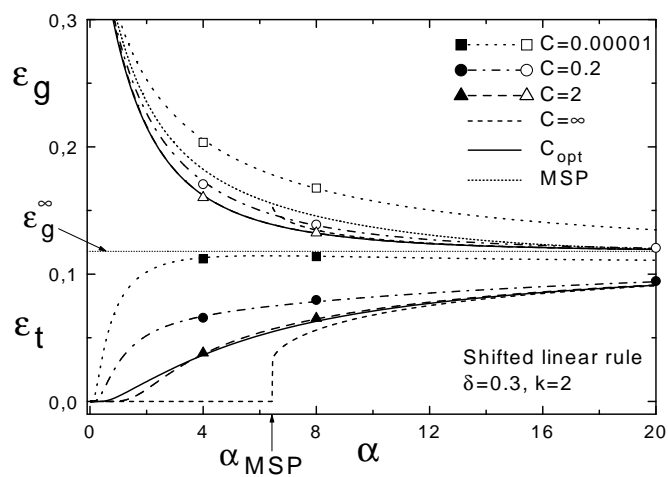,height=7 cm}}
\caption{Shifted linear rule. Same as the preceding
figure, with an exponent $k=2$ in the cost
function. Simulations results correspond to $N=100$.}
\label{fig.4}
\end{figure}

\pagebreak

\begin{figure}
\centerline{\psfig{figure=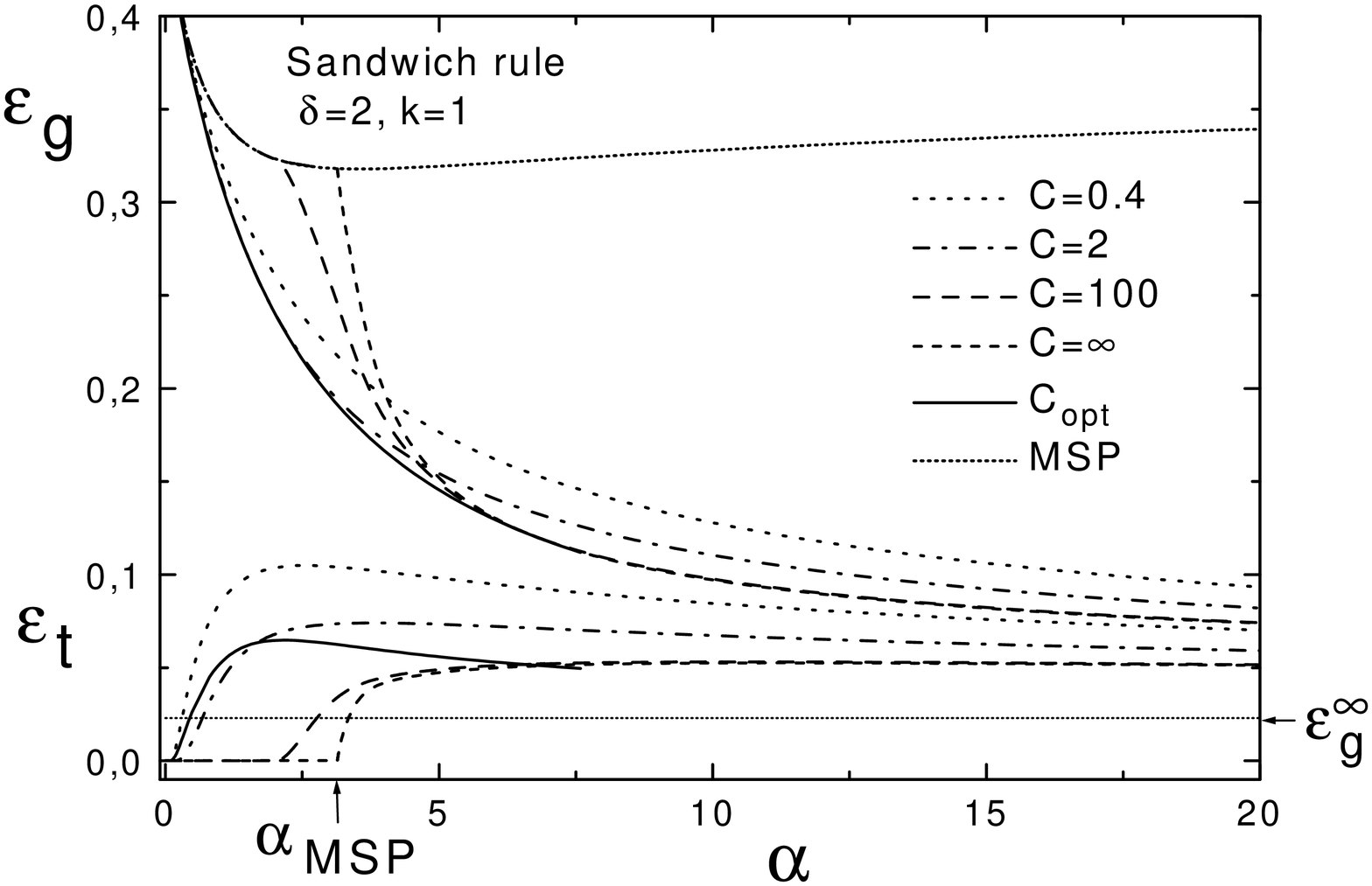,height=7 cm}}
\caption{Sandwich rule. SMC's learning
curves corresponding to an exponent $k=1$ in the cost
function, for different values of the hyperparameter
$C$. Asymptotically, $\epsilon_g^\infty=0.023$.}
\label{fig.5}
\end{figure}

\pagebreak

\begin{figure}
\centerline{\psfig{figure=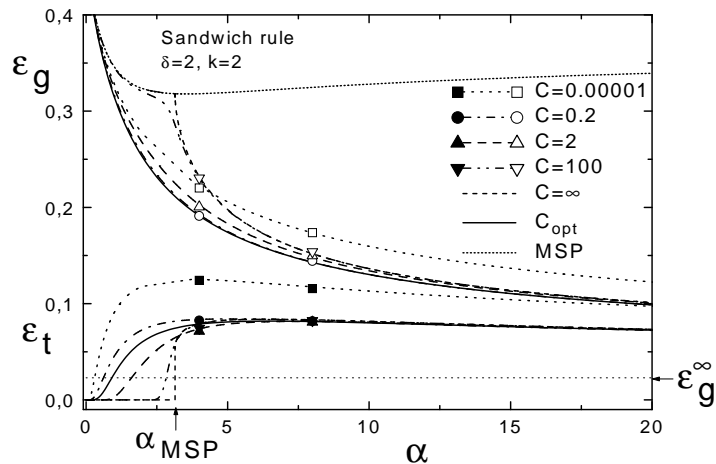,height=7 cm}}
\caption{Sandwich rule. SMC's learning
curves corresponding to an exponent $k=2$ in the cost
function, for different values of the hyperparameter
$C$. Symbols correspond to results of computer
simulations with $N=100$. Asymptotically, $\epsilon_g^\infty=0.023$.}
\label{fig.6}
\end{figure}

\pagebreak

\begin{figure}
\centerline{\psfig{figure=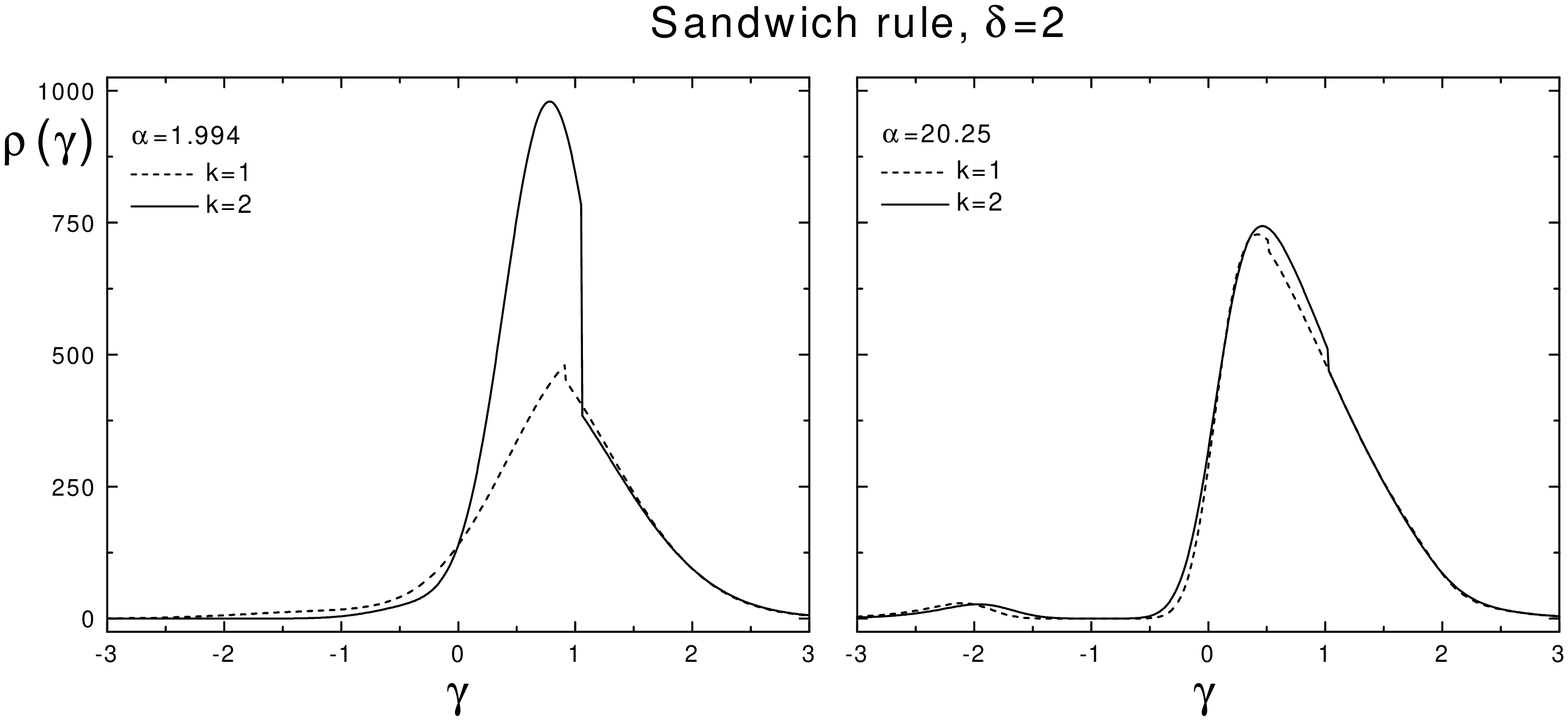,height=7 cm}}
\caption{Sandwich rule. Distribution of stabilities of the SMC
for two different training set sizes $\alpha$, obtained with $C=2$
in the cost function.}
\label{fig.7}
\end{figure}

\pagebreak

\begin{figure}
\centerline{\psfig{figure=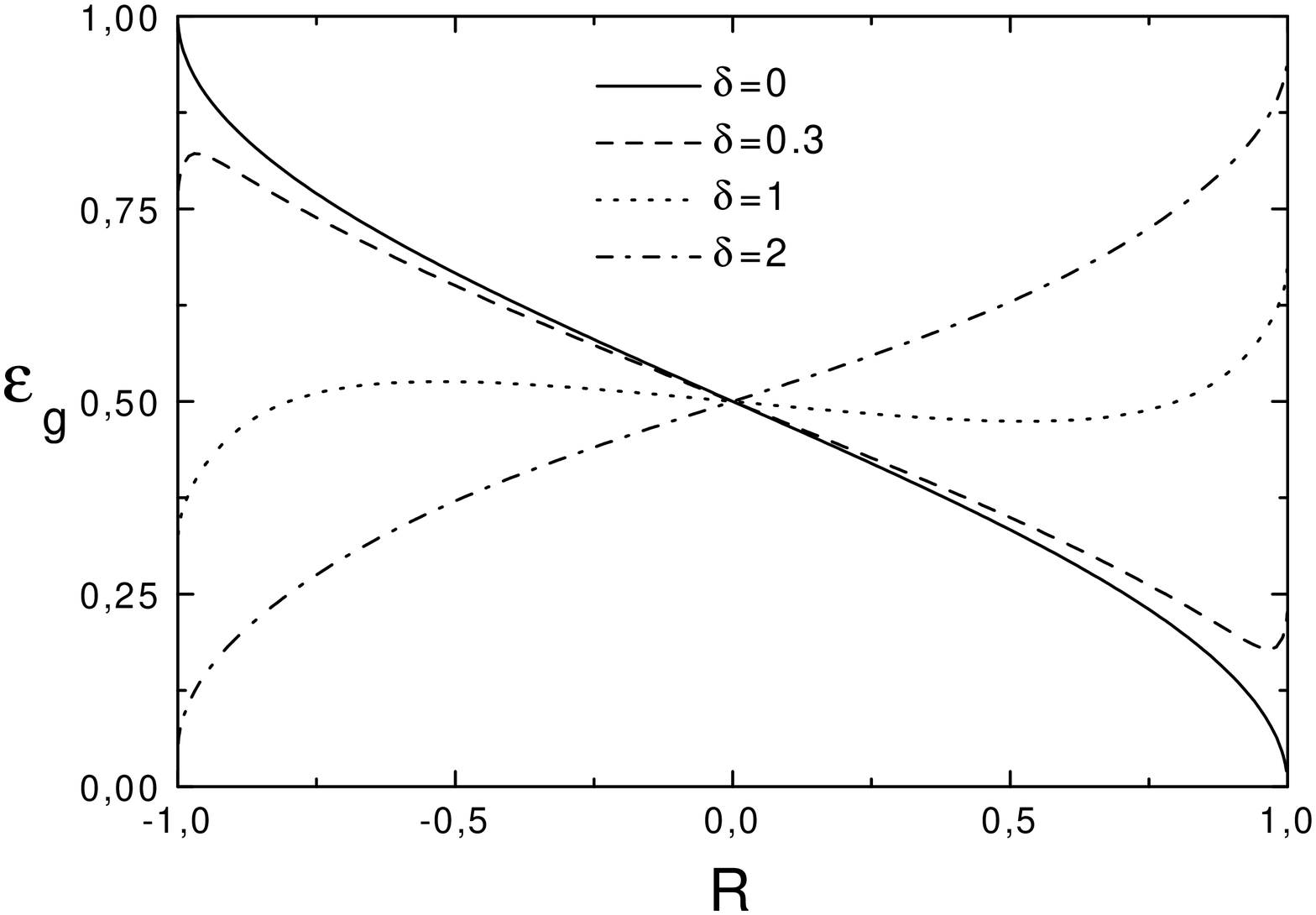,height=7 cm}}
\caption{Reverse wedge rule. Generalization error as a function
of the normalized overlap $R$ between the teacher's and the
student's weight vectors, for different wedge widths $\delta$.}
\label{fig.8}
\end{figure}

\pagebreak

\begin{figure}
\centerline{\psfig{figure=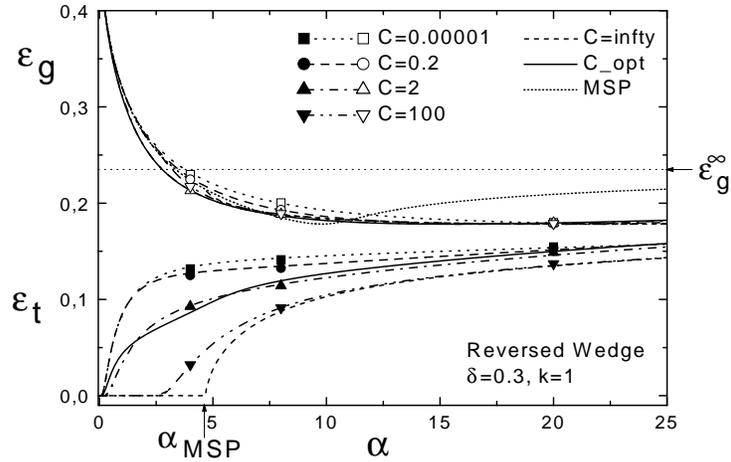,height=7 cm}}
\caption{Reverse wedge rule with $\delta=0.3$. SMC's learning
curves corresponding to an exponent $k=1$ in the cost
function. The optimal value of the generalization error is
$\epsilon_g^{opt}=0.178$, but the SMC converges
asymptotically to $\epsilon_g^\infty=0.235$. Simulation
results correspond to $N=100$.}
\label{fig.9}
\end{figure}

\pagebreak

\begin{figure}
\centerline{\psfig{figure=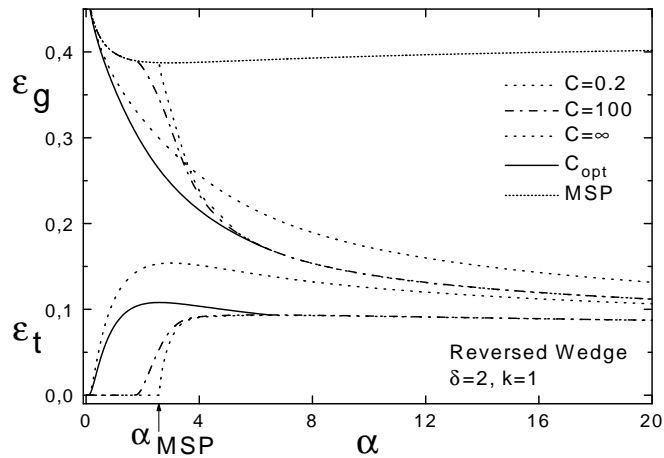,height=7 cm}}
\caption{Reverse wedge rule with $\delta=2$. Learning curves
obtained with different values of the hyperparameter $C$,
with $k=1$ in the cost function. Asymptotically, $\epsilon_g^\infty=0.0455$.
}
\label{fig.10}
\end{figure}

\pagebreak

\begin{figure}
\centerline{\psfig{figure=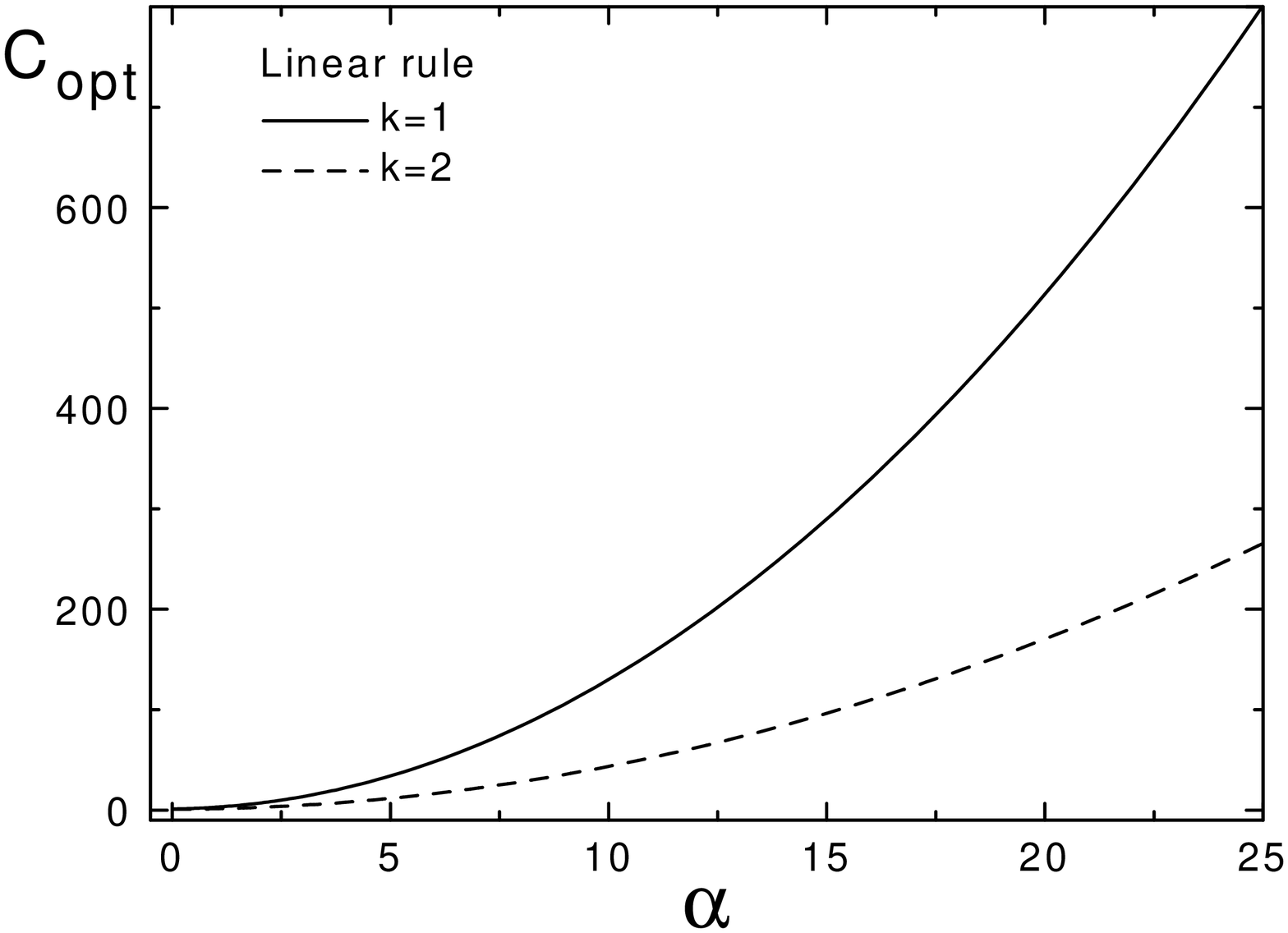,height=7 cm}}
\caption{Linear rule. Optimal values of the hyperparameter $C_{opt}$.}
\label{fig.11}
\end{figure}

\pagebreak

\begin{figure}
\centerline{\psfig{figure=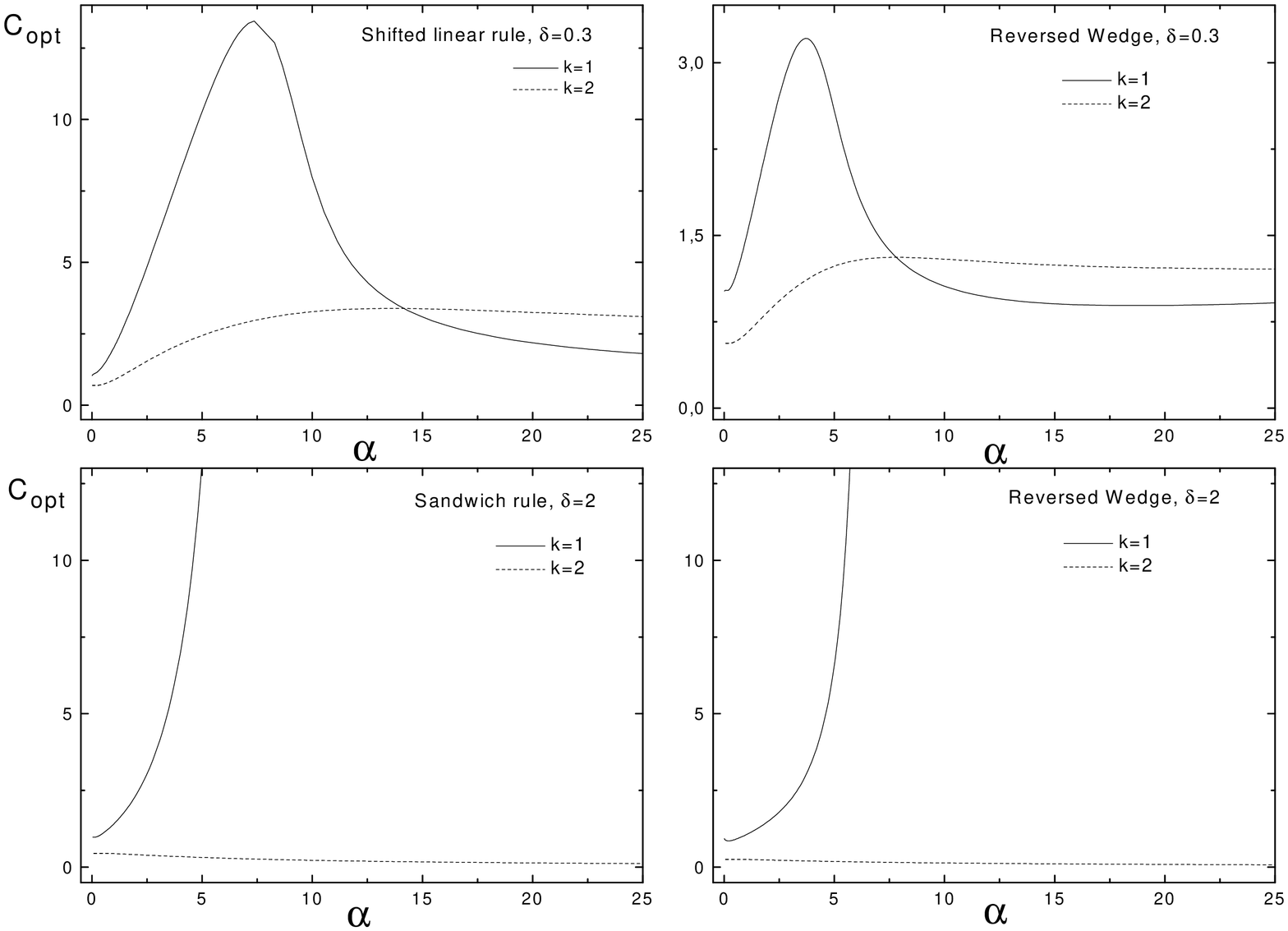,height=7 cm}}
\caption{Optimal values of the hyperparameter $C_{opt}$ for
unrealizable rules.}
\label{fig.11}
\end{figure}


\begin{thebibliography}{99}
\bibitem{foot0} Here, as in the rest of the paper, we call zeros the points where the function changes sign, i.e. they are zeros of the function but {\it not} of its derivative.
\bibitem{foot1} This is true only if the patterns are in general position (that is, if all the patterns in every subset of $N$ patterns are linearly independent)~\cite{BuCr}. For the gaussian pattern distribution we are considering, the probability that $M$ points do not be in general position vanishes in the thermodynamic limit.
\end{thebibliography}
\end{document}